\documentclass[11pt,a4paper]{article}
\pdfoutput=1
\usepackage{jheppubme}
\usepackage{amsfonts}
\usepackage{bbm}
\usepackage{graphicx}
\usepackage{caption}
\usepackage{subcaption}
\usepackage[normalem]{ulem}
\usepackage{amsmath,amssymb,epsfig}
\usepackage{xspace}
\usepackage{comment}
\usepackage{url}

\def\one{{\,\hbox{1\kern-.8mm l}}}
\newcommand{\Dslash}{\not{\hbox{\kern-4pt $D$}}}
\newcommand{\pdslash}{\not{\hbox{\kern-2pt $\partial$}}}
 
 \newcommand{\cO}{\mathcal{O}}

\newcommand{\Comment}[1]{{}}

\def\IZ{{\mathbb Z}}

\newcommand{\bc}{\begin{center}}
\newcommand{\ec}{\end{center}}
\newcommand{\ba}{\begin{array}}
\newcommand{\ea}{\end{array}}
\newcommand{\beq}{\begin{equation}}
\newcommand{\eeq}{\end{equation}}
\newcommand{\bea}{\begin{eqnarray}}
\newcommand{\eea}{\end{eqnarray}}
\newcommand{\bmx}{\begin{pmatrix}}
\newcommand{\emx}{\end{pmatrix}}
\newcommand{\nn}{\nonumber}

\newcommand{\be}{\begin{equation}}
\newcommand{\ee}{\end{equation}}

\newcommand{\g}{\gamma}

\newcommand{\p}{\phi}

\newcommand{\del}{\partial}
\newcommand{\half}{{\frac{1}{2}\,}}
\newcommand{\tr}{{\rm tr}}

\newcommand{\tpsi}{{\widetilde \psi}}

\newcommand{\eref}[1]{Equation~(\ref{#1})}

\newcommand{\valpha}{{\vec\alpha}}
\newcommand{\vbeta}{{\vec\beta}}

\newcommand{\hg}{{\hat{g}}}

\newcommand{\hC}{{\widehat{C}}}

\newcommand{\Renyi}{R\'enyi\xspace}

\def\IC{\mathbb{C}}

\def\IZ{\mathbb{Z}}

\newcommand\sfrac[2]{{\textstyle\frac{#1}{#2}}}

\newcommand\shalf{{\textstyle\frac12}}

\normalsize

\def\a{\alpha}
\def\b{\beta}
 \def\g{\gamma}

  \def\w{\omega}

\def\th{\theta}

\def\t{\tau}

\def\O{\Omega}  \def\w{\omega}

\def\p{\partial}
\def\be{\begin{equation}}
\def\ee{\end{equation}}
\def\bea{\begin{eqnarray}}
\def\eea{\end{eqnarray}}
\def\ba{\begin{align}}
\def\ea{\end{align}}

\newcommand{\bem}{\begin{pmatrix}}
\newcommand{\eem}{\end{pmatrix}}

\def\={\;  = \;}
\def\+{\, + \,}

\def\bar{\overline}

\def\rt2{\sqrt{2}}

\newcommand{\ndt}{\noindent}

\newcommand{\nth}{$n^\text{th}$\xspace}

\newcommand{\e}{{\bf e}}

\newcommand\Thth{$\Theta$\,-\,$\theta$~}
\newcommand\ve{\varepsilon}

\newcommand{\Chat}{\widehat{C}}

\title{Fermions on replica geometries and the $\Theta$\,-\,$\theta$ relation}

\author{Sunil Mukhi$^a$}
\author{and Sameer Murthy$^b$}

\affiliation{$^a$ Indian Institute of Science Education and Research,\\
Homi Bhabha Rd, Pashan, Pune 411 008, India}
\affiliation{$^b$ Department of Mathematics, King's College London,\\
The Strand, London WC2R 2LS, U.K}

\emailAdd{sunil.mukhi@gmail.com}
\emailAdd{sameer.murthy@kcl.ac.uk}

\abstract{In arXiv:1706:09426 we conjectured and provided evidence for an identity between Siegel $\Theta$-constants 
for special Riemann surfaces of genus $n$ and products of Jacobi $\theta$-functions. This arises by comparing two 
different ways of computing the \nth \Renyi entropy of free fermions at finite temperature. Here we show that for~$n=2$ 
the identity is a consequence of an old result due to Fay for doubly branched Riemann surfaces. For~$n>2$ we 
provide a detailed matching of certain zeros on both sides of the identity. This amounts to an elementary proof of the 
identity for $n=2$, while for $n\ge 3$ it gives new evidence for it. 
We explain why the existence of additional zeros renders the general proof difficult.}


\begin{document}

\maketitle

\section{Introduction}

The computation of \Renyi entanglement entropies for quantum field theories is commonly performed 
using the replica trick in the path integral formalism. This requires us to change the underlying space 
on which the field theory is defined to a ``replica space'' which contains $n$ copies of the original one 
glued together in a certain way. This approach has been successfully applied to the computation of 
the \Renyi entropy for a single interval in arbitrary 2d conformal field theories as long as either the 
temperature is zero (non-compact Euclidean time direction) or the space is infinite (non-compact spatial 
direction), or both. The answer is a universal expression that does not depend on any details of the CFT 
other than the central charge. The replica trick has also been applied to the computation of multi-interval 
\Renyi entropies \cite{Calabrese:2009qy,Calabrese:2009ez,Calabrese:2010he}, to entanglement negativity 
\cite{Calabrese:2012ew,Calabrese:2014yza}, and to single-interval \Renyi entropy on a finite space at 
finite temperature~\cite{Azeyanagi:2007bj,Datta:2013hba,Cardy:2014jwa,Chen:2015cna}. In  these cases, 
the result is not universal and depends on details of the CFT beyond the central charge. 

Typically the difference between the cases that give universal answers and those that do not is the nature 
of the Riemann surface involved. For the simplest case of a single interval on a plane, one can uniformise 
the replica surface back to the original one, whereupon the computation is relatively simple. The same is 
true for a single interval on a cylinder (which can be mapped to a punctured plane and one proceeds from 
there). In these cases, the computation boils down to a product of two-point functions of twist fields, which 
on the plane are completely determined by their conformal dimensions. These dimensions in turn are 
given by the uniformising map. However when we have  finite temperature {\em and} interval size, the 
space is a torus and the replica space is a genus-$n$ surface. Similarly when we have $n>1$ cuts on a 
plane, the replica space has genus $n-1>0$. These spaces cannot be uniformised back to the original one. 
For free field theories, however, one can still attempt to use twist fields. When the original space is a torus 
with one cut, one needs their two-point functions on a torus. When the original space is a plane with 
multiple cuts, one needs their higher-point correlation functions. These objects are generally computable 
in free boson and fermion theories. 

In particular, the \Renyi entropy at finite size and temperature has been computed in 
\cite{Azeyanagi:2007bj,Datta:2013hba,Chen:2015cna}. For bosons, the result in \cite{Chen:2015cna}
is consistent and passes all reasonable tests including the thermal entropy relation and modular invariance. 
However, as originally pointed out in \cite{Lokhande:2015zma}, the free fermion result of~\cite{Azeyanagi:2007bj} 
does not simultaneously pass both these tests. If one chooses to restrict to a single fermion spin structure 
then the answer obeys the thermal entropy relation but is not modular invariant. On the other hand, if one 
chooses to sum over spin structures then the answer is modular invariant but does not obey the thermal 
entropy relation. This puzzle was addressed in \cite{Mukhi:2017rex} where it was first argued that the 
twist-field computation of~\cite{Azeyanagi:2007bj}, for a fixed spin structure, is equivalent to computing the 
higher-genus partition function of free fermions for a ``diagonal'' spin structure on a genus-$n$ Riemann surface. 
This is what one would expect, but it implies a very non-trivial identity between genus-$n$ Siegel $\Theta$-constants 
on replica surfaces and genus-1 $\theta$-functions, which we call the $\Theta$\,-\,$\theta$ relation. A precise 
statement of this identity and considerable evidence for it were provided in \cite{Mukhi:2017rex}. The resolution 
of the puzzle is then as follows: the modular-invariant higher-genus partition function for free fermions requires 
a sum over all $2^{n-1}(2^n+1)$ even spin structures on the genus-$n$ surface, including the non-diagonal 
ones. However no twist-field computation is known that reproduces the non-diagonal spin structures. 
Therefore, though the twist-field computation is correct for diagonal spin structures, it is inadequate to find 
the contribution of non-diagonal ones. As a result, the only correct way to compute free fermion \Renyi 
entropies at finite size and temperature is to use the higher-genus approach. 

The above resolution left three interesting questions open. To explain these, let us first set up the problem. 
We have a free fermion CFT on a spatial region of size $L$ and at a finite temperature $T=\beta^{-1}$. 
By standard methods the corresponding path integral is defined on a torus whose horizontal axis is unity 
and the ratio $\tau=\frac{\beta}{L}$, after complexification, describes the other axis of the torus. The entanglement 
of interest is between the interior of a straight line segment between the points $z_1$ and $z_2$ on this torus, 
whose separation is denoted $z_{12}=z_2-z_1$ and allowed to be complex. Computations of \Renyi entropy 
depend only on the two complex parameters $\tau,z_{12}$. Now, one of the open problems was whether a 
rigorous proof can be found for the conjectured identity (Equation~(2.24) of~\cite{Mukhi:2017rex}) between a higher-genus 
Siegel $\Theta$-constant for a diagonal spin structure, evaluated on a special replica surface, and a product of 
genus-1 $\theta$-functions with the same spin structure. Another was whether the twist field prescription can be modified in some way to 
compute the contribution of non-diagonal spin structures. A third question, only briefly addressed in \cite{Mukhi:2017rex}, 
concerns the periodicity of the \Renyi entropy contribution for a fixed diagonal spin structure. These contributions
 to the \Renyi entropy do not have periodicity in the interval length of the original torus ($z_{12}\to z_{12}+1, z_{12}+\tau$)
 but rather, are invariant under $n$-fold shifts for odd~$n$ and $2n$-fold shifts for even $n$. A related point is that 
 these contributions have zeros for values of $z_{12}$ that lie outside the original torus but inside the larger 
 fundamental region for the $n$-fold/$2n$-fold shifts. 

In this note, we address all three points above. 
After a brief review of the results of \cite{Mukhi:2017rex}, we argue that the twist-field prescription cannot be 
used for non-diagonal spin structures, for the simple reason that replica symmetry does not hold in this situation. 
Next we provide a rigorous proof of the conjectured identity of \cite{Mukhi:2017rex} for the case $n=2$ 
(corresponding to the second \Renyi). In fact this identity is equivalent to a mathematical result of Fay \cite{Fay:book} 
on doubly-branched covers, for which we also provide an elementary proof which relies on information about the 
zeros of the $\theta$-functions that appear in the identity. For $n>2$ the corresponding identity does not seem to 
be known in the mathematics literature. We analyse in some detail the pattern of zeros for the twist-field calculation 
at $n>2$, which is significantly different from that at $n=2$, and use it to propose some directions in which a 
rigorous proof for $n>2$ might be found. 
We are able to match some of the zeros on the two sides, but for other zeros of the right-hand side 
we are unable to compute the left-hand side and verify whether it vanishes.
Finally we discuss some interesting open questions to which these techniques could be applied. 

The plan of the paper is as follows. In Section~\ref{reviewof} we start by reviewing the conjecture of~\cite{Mukhi:2017rex} 
for what we call ``diagonal'' spin structures, and then go on to argue that analogous relations do not exist for 
non-diagonal spin structures. In Section~\ref{secondren} we give a proof of the~\Thth identity for~$n=2$ using 
a result of Fay. In Section~\ref{sec:zeros} we discuss the periodicity relations and zeros of the two sides of the 
identity, leading to additional evidence for it. This includes an elementary proof of the conjecture for~$n=2$. 
We conclude in Section~\ref{sec:conclusion} with remarks on the geometric interpretation of our identity.

\section{\Renyi entropy computations and the conjectured $\Theta-\theta$ relation \label{reviewof}}

Given a reduced density matrix $\rho_A$ obtained by tracing out the degrees of freedom outside the 
entangling interval, the $n^{\text{th}}$ \Renyi entropy is defined as: 
\be
S_n(z_{12},\tau)\=\frac{1}{1-n}\log\tr\, \rho_A^n
\ee 
It has been known for some time that this can be computed in terms of a ``replica partition function'' 
on a genus-$n$ Riemann surface:
\be
S_n\=\log Z^{(n)}
\ee
For free fermions/bosons, the partition function on generic Riemann surfaces is known, so in principle 
one just needs to specialise the answer to the replica surface. 

In particular, the partition function of a modular-invariant free Majorana fermion theory on a compact 
Riemann surface of genus $n$ is \cite{AlvarezGaume:1986es,Dijkgraaf:1987vp}:
\be
Z^{(n)}_\text{higher-genus}\biggl[\begin{matrix}\valpha \\ \vbeta\end{matrix}\biggr](\Omega)
\=\frac{1}{2^n}|{\cal C}| \,\sum_{\valpha,\beta}
\bigg|\Theta\bigg[\begin{matrix}\valpha \\ \vbeta\end{matrix}\bigg](0|\Omega)\bigg| \,,
\label{higher-genus-gen}
\ee
where $\Theta$ is the genus-$n$ Siegel theta-function with characteristics $\valpha,\vbeta$:
\be
\Theta\biggl[\begin{matrix}\valpha \\ \vbeta\end{matrix}\biggr](0|\Omega)
\= \sum_{m_{i} \in \IZ + \a_{i}} 
\exp\Bigl(\pi i  \sum_{i,j=1}^2 m_i \, \O_{ij} \, m_{j}+ 2 \pi i \sum_{i=1}^2 m_i \, \b_{i} \Bigr) \, .
\ee
In \eref{higher-genus-gen}, the quantity ${\cal C}$ is a factor related to the determinant of an 
anti-holomorphic differential operator on the surface (see for example Eq.(5.13) of~\cite{AlvarezGaume:1986es}). 

If we wish to compute \eref{higher-genus-gen} on a replica surface, we must insert the appropriate 
period matrix $\Omega$ for this surface. This is defined in terms of a set of $n$ cut differentials:
\be
\omega_k(z,z_{12},\tau) \,:= \, \frac{\theta_1(z|\tau)}{
\theta_1\Big(z+\frac{k}{n}z_{12}\Big|\tau\Big)^{1-\frac{k}{n}}\theta_1\Big(z-(1-\frac{k}{n})z_{12}
\Big|\tau\Big)^{\frac{k}{n}}} \,,
\label{cutdiff}
\ee
with $k=0,1,\cdots,n-1$\footnote{For the odd Jacobi theta function we will often use the standard 
notation~$\theta_1(z|\tau)=-\theta\left[\begin{matrix}\half \\ \half\end{matrix}\right](z|\tau)$ to reduce 
clutter in our formulae.}. From these differentials we construct a set of $n$ quantities:
\be
C_k(z_{12},\tau)\, :=\, \frac{\int_0^\tau\omega_k(z,z_{12},\tau)\,dz}{\int_0^1\omega_k(z,z_{12},\tau)\,dz} \,,
\ee
and the period matrix $\Omega$ is then given by:
\be
\Omega_{ab}(z_{12},\tau)\=\sum_{k=0}^{n-1} e^{\frac{2\pi i (a-b)k}{n}}C_k(z_{12},\tau)\,.
\ee
More details, including original references, can be found in \cite{Mukhi:2017rex}.

The sum over characteristics in the above equation is a sum over fermion boundary conditions over 
closed cycles, or spin structures. Thus we can write $\valpha=(\alpha_1,\cdots,\alpha_n)$ and 
$\vbeta=(\beta_1,\cdots,\beta_n)$ where each $\alpha_i,\beta_j$ independently takes values 0 or $\half$. 
The value 0 indicates an anti-periodic fermion boundary condition around the $A$-cycle (for $\alpha_i$) or 
the $B$-cycle (for $\beta_i$), while the value $\half$ indicates a periodic boundary condition around 
the same cycle. Diagonal spin structures are those of the form $(\alpha,\cdots,\alpha)$ and $(\beta,\cdots,\beta)$ 
and will be denoted by $\valpha_{\rm diag}$ and $\vbeta_{\rm diag}$ in what follows. It is only for diagonal spin 
structures that the fermion has the same boundary condition around every $A$-cycle, and similarly for 
the $B$-cycles. Notice that in the $3n-3$ complex-dimensional parameter space of compact genus-$n$ 
surfaces for $n\ge 2$, this set of period matrices is a subfamily of dimension 2.

\subsection{Twist fields and the conjecture}

Restricting to replica surfaces, it is possible to write a more explicit form for the replica partition function 
in which the spin-structure-independent prefactor ${\cal C}$ is made precise. The result, as derived in \cite{Mukhi:2017rex}, is:
\be
Z^{(n)}_\text{higher-genus}\=
\left|\frac{\theta_1'(0|\tau)}{\theta_1(z_{12}|\tau)}\right|^{\frac1{12}(n-\frac{1}{n})}\frac{1}{|\eta(\tau)|^{n}}
\frac{\sum_{\valpha,\vbeta}
\Big|\Theta\bigg[\begin{matrix}\valpha \\ \vbeta\end{matrix} \bigg](0|\Omega)\Big|}
{\sqrt{\prod_{k=1}^{n-1}
{\int_0^1\omega_k(z,z_{12},\tau)\,dz}}}
\,.
\label{higher-genus}
\ee

The twist-field computation of a replica partition function attempts to reproduce the same result by computing 
a correlation function on the original torus. As discussed in \cite{Lokhande:2015zma,Mukhi:2017rex}, this has been 
only partially successful for free fermions. Indeed, this method was initially applied to compute the contribution 
to the replica partition function for a fixed spin structure $(\alpha,\beta)$ on the original torus, and yielded the 
result~\cite{Azeyanagi:2007bj}:
\be
Z^{(n)}_\text{twist-field} \=
\left|\frac{\theta_1'(0|\tau)}{\theta_1(z_{12}|\tau)}\right|^{\frac{1}{12}(n-\frac{1}{n})}\prod_{k=-{\frac{n-1}{2}}}^{\frac{n-1}{2}}\frac{\left|\theta\bigg[\begin{matrix}\alpha\\ \beta\end{matrix}\bigg]\left(\frac{k}{n}z_{12}\Big|\tau\right)\right|}{|\eta(\tau)|} \,,
\label{twist}
\ee
This is supposed to represent an alternate approach to the computation of the free-fermion \Renyi entropy. 
But as it stands, this depends on a pair $(\alpha,\beta)$ specifying torus spin structures, and cannot be 
compared with \eref{higher-genus} where $2^{2n}$ spin structures have been summed over. What we should
do is to compare the twist-field result in \eref{twist} with the contribution to \eref{higher-genus} from a fixed, 
diagonal spin-structure $(\valpha_{\rm diag},\vbeta_{\rm diag})$ as defined above. A sufficient condition for 
these two to agree is the following non-trivial mathematical identity between higher-genus $\Theta$-constants 
(evaluated on the period matrix of the replica surface) and genus-1 $\theta$-functions, which was conjectured in \cite{Mukhi:2017rex}:
\be 
\chi_{g}(\tau,z_{12};\a,\b) \= \chi_{t}(\tau,z_{12};\a,\b)\,,
\label{MainIdentity}
\ee
with
\be 
\label{defchig}
\chi_{g}(\tau,z_{12};\a,\b) \= \frac{\Theta\bigg[\begin{matrix}\valpha_\text{diag}\\ \vbeta_\text{diag}\end{matrix}\bigg](0|\Omega)}{\sqrt{\prod_{k=1}^{n-1}
{\int_0^1\omega_k(z,z_{12},\tau)\,dz}}} \,,
\ee
\be 
\label{defchit}
\chi_{t}(\tau,z_{12};\a,\b) \= 
\prod_{k=-{\frac{n-1}{2}}}^{\frac{n-1}{2}}\theta\bigg[\begin{matrix}\alpha\\ \beta\end{matrix}\bigg]\left(\frac{k}{n} z_{12}\Big|\tau\right)\,.
\ee
We will refer to~\eref{MainIdentity} as the \Thth identity.
The above conjecture is slightly stronger than what is required for the partition functions coming  
from~\eref{higher-genus} and \eref{twist} to be 
equal, in that it equates {\em holomorphic} functions of $(z_{12}, \tau)$. Several pieces of evidence were 
provided for this identity in \cite{Mukhi:2017rex}. In particular it was shown that both sides transform in the 
same way (as weak Jacobi forms) under modular transformations of the torus, and that they have the same 
periodicities in the variable~$z_{12}$. These results will be useful in the following. 

It is evident that restricting the sum in \eref{higher-genus-gen} to diagonal spin structures does not lead to an 
answer that is invariant under modular transformations. An easy way to see this is that such a sum contains 
just four terms corresponding to $(\alpha,\beta)=(0,0), (0,\half),(\half,0),(\half,\half)$, but a generic global 
diffeomorphism of a genus-$n$ Riemann surface---such as cutting, twisting and re-joining a single handle---will 
change the boundary conditions to correspond to a non-diagonal spin structure. We have seen that if our 
conjecture is true then twist fields can reproduce the higher-genus result for diagonal spin structures, but this 
motivates us to ask what is the situation for non-diagonal spin structures and whether there is a corresponding identity for them. 

\subsection{Non-diagonal spin structures}

We have seen that the spin-structure-dependent part of the partition function of a higher-genus Riemann 
surface is given (for Majorana fermions) by a sum:
\be
\sum_{\valpha,\vbeta}\left|\Theta\bigg[\begin{matrix}
\valpha\\ \vbeta
\end{matrix}\bigg](0|\Omega)\right|
\ee
where $\Omega$ is the period matrix of the Riemann surface, which is a complicated function of $(\tau,z_{12})$. 
Our conjecture implies that twist fields can reproduce the above quantity in the case where the higher-genus spin 
structure is diagonal, of the form $\valpha_{\rm diag}$, $\vbeta_{\rm diag}$. One is therefore naturally led to ask 
whether some modification of these twist fields can be employed to reproduce {\em non-diagonal} spin structures 
of the higher-genus surface.

Unfortunately, as we now argue, this is not possible. The replica method works due to the $Z_n$ replica 
symmetry, but this symmetry is broken precisely by the boundary conditions that correspond to non-diagonal 
spin structures. To be more precise, on the higher-genus surface one can still define operators $\sigma$ 
located at the end-points of the entangling interval that take the physical fields from one replica to the next:
\be
\sigma(z)\psi_j(w) \, \sim \, \psi_{j+1}(w)
\ee 
where $\psi_j,j=1,\cdots,n$ is the free fermion associated to the $i$th replica. However the replica method 
really becomes useful when one takes linear combinations of the replicated physical fields and reduces 
the genus-$n$ problem to a problem in genus-1 with multiple fields, each acquiring a different phase upon 
encircling the end-points of the cut:
\be
\tpsi_k\=\sum_{j=1}^n e^{2\pi ijk/n}\psi_j \,,\qquad \sigma(z)\tpsi_k(w) \,\sim \, e^{2\pi ik/n}\tpsi_k(w)
\ee
It is evident that if all the $\psi_i$ do not have the same boundary condition around the corresponding cycle 
of the replica surface, then the $\tpsi_k$ do not simply pick up an overall phase when acted on by the twist 
field. Instead they are mixed by the monodromies. In short, the ``diagonalising'' process fails with non-diagonal 
spin structures, and one cannot convert replica fermions on a genus-$n$ surface into free fermions on a single 
torus. An inevitable conclusion is that the \Renyi entropy for modular-invariant free fermion systems cannot 
be computed using the twist-field method.

\section{Second \Renyi entropy: a proof of the identity}

\label{secondren}

In this section we focus on the second \Renyi entropy, i.e.~the case when the replica surface has genus $n=2$. 
This surface is a doubly branched cover of the torus. Such covers have been studied quite intensively and 
we will see that useful formulae exist in the mathematical literature which enable us to prove the \Thth identity in this case.

We start by specialising the notation of Section \ref{reviewof} to the case of genus 2. One easily sees that the cut differentials 
in this case are
\be
\omega_1(z,z_{12},\tau)   \= \, \frac{\theta_1(z|\tau)}{
\sqrt{\theta_1\Big(z+\half z_{12}\Big|\tau\Big)\theta_1\Big(z-\half z_{12}\Big|\tau\Big)}} \,,
\ee
and that the period matrix is
\be
\begin{split}
\Omega &\= \half \begin{pmatrix} \tau+C_1 &\tau-C_1\\ \tau-C_1&\tau+C_1\end{pmatrix} \, , \\[2mm]
C_1(z_{12},\tau) &\=\frac{\int_0^\tau\omega_1(z,z_{12},\tau)\,dz}{\int_0^1\omega_1(z,z_{12},\tau)\,dz} \, .
\label{pergentwo}
\end{split}
\ee
%\be
%\begin{split}
%\omega_1(z,z_{12},\tau)   &\= \, \frac{\theta_1(z|\tau)}{
%\sqrt{\theta_1\Big(z+\half z_{12}\Big|\tau\Big)\theta_1\Big(z-\half z_{12}\Big|\tau\Big)}} \,, \\
%\Omega &\= \half \begin{pmatrix} \tau+C_1 &\tau-C_1\\ \tau-C_1&\tau+C_1\end{pmatrix} \, , \\[2mm]
%C_1(z_{12},\tau) &\=\frac{\int_0^\tau\omega_1(z,z_{12},\tau)\,dz}{\int_0^1\omega_1(z,z_{12},\tau)\,dz} \, .
%\label{pergentwo}
%\end{split}
%\ee
Our conjectured identity \eref{MainIdentity} becomes, in this case,
\be
\frac{\Theta\bigg[\begin{matrix}\alpha &\alpha\\ \beta & \beta\end{matrix}\bigg](0|\Omega)}{\sqrt{\int_0^1\omega_1(z,z_{12},\tau)\,dz}} \=
\theta\bigg[\begin{matrix}\alpha\\ \beta\end{matrix}\bigg]\left(\frac{z_{12}}{4}\Big|\tau\right)
\theta\bigg[\begin{matrix}\alpha\\ \beta\end{matrix}\bigg]\left(-\frac{z_{12}}{4}\Big|\tau\right)\,,
\label{gentwoid}
\ee
In \cite{Mukhi:2017rex} we showed that both sides have the same periodicity under shifts $z_{12}\to z_{12}+4$, 
and provided additional evidence by expanding both sides in powers of $z_{12}$. In this section we 
prove \eref{gentwoid} using an old result due to Fay \cite{Fay:book}. 
To start with, let us review some basic features of ramified double coverings. The idea is to describe the 
simplest class of Riemann surfaces $\hC$ with non-trivial automorphism group ${\rm Aut}(\hC)$ and a ramified projection mapping:
\be
\hC\to \hC/{\rm Aut}(\hC)
\ee
This class consists of Riemann surfaces admitting a conformal involution with fixed points. 

Although the theory applies to such Riemann surfaces of any genus, 
we will specialise it to the case where the covering space has genus $\hg=2$ and 
the base has genus $g=1$. Thus, let $\pi: \hC \to C$ be a ramified double covering of genus 2 of a torus 
with 2 branch points at $Q_1,Q_2$. Let $\phi:\hC\to \hC$ be the conformal automorphism that exchanges the two tori 
in the covering surface, with fixed points at $Q_1,Q_2$. A canonical homology basis is then:
\be
A_1\,,B_1\,,A_1'\,,B_1'\,,
\ee
where $A_1,B_1$ is a canonical homology basis for the torus $C$, and
\be
A_1'\=-\phi(A_1) \,,\qquad B_1'\=-\phi(B_1)
\label{canhom}
\ee
Correspondingly there are normalised holomorphic differentials $u_1,u_1'$ where
\be
u_1'(x')\=-u_1(x) \,,
\label{normdiff}
\ee
and $x'=\phi(x)$ is the conjugate point of $x\in\hC$ under the automorphism.

An alternate basis for the holomorphic differentials is given by
\be
v_1\=u_1-u_1' \,,\qquad w_1\=u_1+u_1' \,.
\ee
Notice that 
\be
v_1(x)\=u_1(x)-u'_1(x)\=u_1(x)+u_1(\phi(x)) \,,
\ee
where we have used $\phi^{-1}=\phi$, i.e.~$\phi$ is of order 2. Clearly $v_1(\phi(x))=v_1(x)$. 
Thus $v_1$ is invariant under the automorphism, and is identified as the (normalised) holomorphic 
differential on the torus $C$. On the other hand,
\be
w_1(x)\=u_1(x)+u_1'(x)\=u_1(x)-u_1(\phi(x))
\ee
is odd under the automorphism. 

The above construction has an interesting geometric interpretation. To every Riemann surface~$C$ 
of genus~$g \ge 1$, one can associate the \emph{Jacobian variety}~$J(C):=\IC^g/(\IZ \, \Omega_g + \IZ \, \mathbbm{1}_g)$. 
The map is given as follows: one chooses an arbitrary point~$P_0 \in C$, and a point~$P \in C$ is then mapped to the 
point~$\int_{P_0}^P \zeta_j \,, j = 1,\cdots, g$ in~$J(C)$,
where~$\zeta_j$ are the~$g$ holomorphic one-forms on the Riemann surface. 
(This map is well-defined because two paths between the two points always differ by a linear combination of cycles on the surface, 
which map to the identity in the quotient that defines~$J$.)
Now, the map~$\pi: \hC \to C$ can be lifted in a canonical manner to a map~$\psi: J(\hC) \to J(C)$.
The map~$\psi$ is actually a homomorphism, and its kernel itself is an abelian variety, known as the \emph{Prym variety}.

In our problem above the Prym variety is a one (complex)-dimensional torus and the holomorphic differential~$w_1$,
called the Prym differential on~$\hC$, is the (normalised) holomorphic differential on this torus.
In \cite{Fay:book} the modular parameters for $\hC$, $C$ and the Prym variety are 
denoted ${\widehat\tau}$, $\tau$, and $\Pi$ respectively. 
In this notation one has the following expressions for these parameters:
\be
\tau \,\equiv \, \int_{B_1}v_1 \,,\qquad \Pi \,\equiv \, \int_{B_1}w_1 \,,
\ee
and
\be
{\widehat\tau} \=\half\begin{pmatrix}
\int_{B_1}u_1 &~ \int_{B_1}u_1' \\[2mm]
\int_{B_1'}u_1 &~ \int_{B_1'}u_1' \\[2mm]
\end{pmatrix} \,.
\ee
Notice that $\tau$ is odd and $\Pi$ is even under the automorphism~$\phi$. This follows from:
\be
\phi(B_1)\=-B_1' \,,\qquad \phi_*(v_1)\=v_1\,,\qquad \phi_*(w_1)\=-w_1 \,,
\ee
where~$\phi_*$ is the induced map on the differentials. 
From the above definitions, it immediately follows that
\be
{\widehat\tau}\=\half\begin{pmatrix}
\tau+\Pi &-\tau+\Pi \\
-\tau+\Pi &\tau+\Pi 
\end{pmatrix} \,.
\label{Omegamat}
\ee

Let us now relate parameters in the notations of Fay and of the present paper. 
In our notation $\tau,{\widehat\tau}$ are denoted $\tau,\Omega$, respectively. Further, we have:
\be
\tau\=\frac{\int_{B_0} \omega_0}{\int_{A_0}\omega_0}\=\int_{B_0}\omega_0^{\text{norm}} \,,
\ee
where
\be
\omega_0^{\text{norm}}\,\equiv \,\frac{\omega_0}{\int_{A_0}\omega_0} \,.
\ee
(These equations are trivial since in practice $\omega_0^{norm}=\omega_0=1$, still they are useful 
in making the correct geometric identification.) Thus we identify
\be
(A_0,B_0)_{\text{us}}\to (A_1,B_1)_{\text{Fay}} \,,
\ee
and
\be
(\omega_0^{norm})_{\text{us}}\=(v_1)_{\text{Fay}} \,.
\ee
If we also identify
\be
(\omega_1^{norm})_{\text{us}}=(w_1)_{\text{Fay}} \,, 
\ee
then we have:
\be
\label{pieqc}
\Pi\=\int_{{B_1}}w_1 \= \int_{{B_0}_{\text{us}}} \omega_1^{\text{norm}}\=\frac{\int_{B_0}\omega_1}{\int_{A_0}\omega_1}\=C_1 \,.
\ee

The minus sign in identifying the period matrices, noted above, appears because of a difference in choice 
of $\phi$ between \cite{Fay:book} and the present work. The former has $u_1'=-\phi_*(u_1)$ while we 
have implicitly chosen $u_1'=\phi_*(u_1)$. Similarly $B_1'=-\phi(B_1)$ in \cite{Fay:book}, while we took it 
as $\phi(B_1)$\footnote{This can be seen in \cite{Mukhi:2017rex} by the fact that $B_0,B_1$ represent 
the ``same'' cycle on the two copies of the torus (i.e. they are mapped to each other by $\phi$ rather 
than $-\phi$).}. This results in the observed sign difference for the off-diagonal elements.

In this context, Proposition~(5.10) of \cite{Fay:book} relates the theta function with modular parameter $2\tau$ to a theta function 
with modular parameter $2C_1$: 
\be
\th\biggl[\begin{matrix}\a \\ \b \end{matrix}\biggr](0|2 C_1) 
\= \Bigl(c\bigl(\frac12(a+b) \bigr) \, c(a) \Bigr)^{\half} \, e^{2 \pi i \a\b} \,
\th\biggl[\begin{matrix}\a \\ \b \end{matrix}\biggr]\Bigl(\frac12 (b-a) | 2 \tau \Bigr) \,, 
\label{Fay.117}
\ee
where~$c(x)$ is a holomorphic section of a line bundle on the genus~2 surface, 
independent of the spin structure and $a,b$ are the end-points of the branch cut.

We can now return to our identity, \eqref{gentwoid}, and manipulate the genus-2 theta-function to bring 
it into a form where the above results can be employed. For a period matrix of the form:
\be
\O \= \half \begin{pmatrix} x+y & x-y  \\ x-y & x+y\end{pmatrix} \,,
\ee
the theta function can be written as:
\be
\Theta\biggl[\begin{matrix}\valpha \\ \vbeta\end{matrix}\biggr](0|\Omega)
\= \sum_{n_{1} = m_{1} + m_{2} \atop n_{2} = m_{1} - m_{2}} 
\e \Bigl(\frac14 \bigl(n_1^{2} \, x + n_{2}^{2} \, y \bigr) +  \frac12 \bigl( n_{1} (\b_{1}+\b_{2})
+ n_{2} (\b_{1}-\b_{2}) \bigr) \Bigr) \,.
\ee
The constraints in the sum are solved by:
\bea
n_{1} \= 2 n_{1}' \,, \; n_{1}' \in \IZ + \frac12(\a_{1}+\a_{2}) \,, \quad 
&& n_{2} \= 2 n_{2}' \,, \; n_{2}' \in \IZ + \frac12(\a_{1}-\a_{2}) \,,  \\
 \text{or} \quad n_{1} \= 2 n_{1}' \,, \; n_{1}' \in \IZ + \frac12(\a_{1}+\a_{2}+1) \,, \quad 
&& n_{2} \= 2 n_{2}' \,, \; n_{2}' \in \IZ + \frac12(\a_{1}-\a_{2}+1) \,.~~~~~~~~
\eea
Writing the theta function in terms of these new variables we obtain:
\bea
\Theta\biggl[\begin{matrix}\valpha \\ \vbeta\end{matrix}\biggr](0|\Omega)
&\= & \sum_{\gamma=0,\half}  \sum_{n_{1}'\in \IZ +\g + (\a_{1}+\a_{2})/2 \atop n_{2}'\in \IZ +\g + (\a_{1}-\a_{2})/2} 
\e \Bigl(\frac12 \bigl(n_1'^{2} \, 2x + n_{2}'^{2} \, 2y \bigr) +  \bigl( n_{1}' (\b_{1}+\b_{2})
+ n_{2}' (\b_{1}-\b_{2}) \bigr) \Bigr) \,, \nn \\
&\= & \sum_{\gamma=0,\half} \th\biggl[\begin{matrix}\g + (\a_{1}+\a_{2})/2 \\ \b_{1}+\b_{2} \end{matrix}\biggr] (0|2x) \;
\th\biggl[\begin{matrix}\g + (\a_{1}-\a_{2})/2 \\ \b_{1}-\b_{2} \end{matrix}\biggr](0|2y) \,.
\eea
For diagonal spin structures, we have~$\a_1=\a_2=\a$ and~$\b_1=\b_2=\b$, so that
\bea
\Theta\biggl[\begin{matrix}\valpha_\text{diag} \\ \vbeta_\text{diag}\end{matrix}\biggr](0|\Omega)
&\= & \sum_{\gamma=0,\half} \th\biggl[\begin{matrix}\g + \a \\ 2\b  \end{matrix}\biggr] (0|2x) \;
\th\biggl[\begin{matrix}\g  \\ 0 \end{matrix}\biggr](0|2y) \,,\\
&\= & \sum_{\gamma=0,\half} e^{4 \pi i (\gamma+\alpha)\beta} \, \th\biggl[\begin{matrix}\g + \a \\ 0  \end{matrix}\biggr] (0|2x) \;
\th\biggl[\begin{matrix}\g  \\ 0 \end{matrix}\biggr](0|2y) \,.
\eea
Applying this identity to our period matrix~\eqref{pergentwo} we obtain:
\be \label{Thth1th2rel}
\Theta\biggl[\begin{matrix}\valpha_\text{diag} \\ \vbeta_\text{diag}\end{matrix}\biggr](0|\Omega)
\=  \sum_{\gamma=0,\half} e^{4 \pi i (\gamma+\alpha)\beta} \, \th\biggl[\begin{matrix}\g + \a \\ 0  \end{matrix}\biggr] (0|2\t) \;
\th\biggl[\begin{matrix}\g  \\ 0 \end{matrix}\biggr](0|2C_1) \,.
\ee
Applying \eref{Fay.117} to the second theta-function on the right-hand side,  and  defining $(b-a) = z_{12}$, we reach the following result:
\be \label{Thth1th2reltau}
\Theta\biggl[\begin{matrix}\valpha_\text{diag} \\ \vbeta_\text{diag}\end{matrix}\biggr](0|\Omega)
 \=  (-1)^{4\a\b} k(\t,z_{12})  \sum_{\gamma=0,\half}  e^{4 \pi i \b \g} \, \th\biggl[\begin{matrix}\g +\a \\ 0 \end{matrix}\biggr] (0|2\t) \;
\th\biggl[\begin{matrix}\g  \\ 0\end{matrix}\biggr] \Bigl(\frac{z_{12}}{2} \Big|2\t \Bigr)  \,,
\ee
where the prefactor~$k(\t,z_{12})=\bigl(c\bigl(\frac{z_{12}}{2} \bigr) \, c(0) \bigr)^{\half}$ depends only on the original torus and the cut length. 
Using the doubling identity for theta functions (\cite{Fay:book}, Equation~4), 
our result~\eqref{Thth1th2reltau} becomes 
\bea \label{Thth1th2reldiag}
\Theta\biggl[\begin{matrix}\valpha_\text{diag} \\ \vbeta_\text{diag}\end{matrix}\biggr](0|\Omega)
& \= &  (-1)^{4\a\b} \,k(\t,z_{12})\, \th\biggl[\begin{matrix}\a \\ \b \end{matrix}\biggr] \Bigl(\frac{z_{12}}{4} \Big|\t \Bigr)   \;
\th\biggl[\begin{matrix} \a \\ \b\end{matrix}\biggr]\Bigl(\frac{z_{12}}{4} \Big|\t \Bigr)   \,, \nn \\
& \= &  \,k(\t,z_{12})\, \th\biggl[\begin{matrix}\a \\ \b \end{matrix}\biggr] \Bigl(\frac{z_{12}}{4} \Big|\t \Bigr)   \;
\th\biggl[\begin{matrix} \a \\ \b\end{matrix}\biggr]\Bigl(-\frac{z_{12}}{4} \Big|\t \Bigr)   \,. 
\eea

Thus one reaches the non-trivial relation between the genus-2~$\Theta$-constant and the genus-1~$\theta$-function.
According to our conjecture, the prefactor~$k$ should be equal to the denominator in the definition~\eqref{defchig} of the 
higher-genus expression~$\chi_g$, which was interpreted in~\cite{Mukhi:2017rex} as the determinant of the 
anti-holomorphic operator~$\bar{\del}$. A simple way to fix this is to look at the modular transformation properties of both 
sides of our main equality and notice that this determinant transforms exactly as the ratio 
$\Theta\biggl[\begin{matrix}\valpha_\text{diag} \\ \vbeta_\text{diag}\end{matrix}\biggr](0|\Omega)\Big/\th\biggl[\begin{matrix}\a \\ 
\b \end{matrix}\biggr] \Bigl(\frac{z_{12}}{4} \Big|\t \Bigr)^2$. This was shown in~\cite{Mukhi:2017rex} and it thus 
proves the equality of our main conjectured identity, up to an overall constant, which was also fixed in~\cite{Mukhi:2017rex} 
to be unity. 

It is worth discussing the non-trivial relation~\eqref{Thth1th2reldiag} a bit more. 
The essential point of the proof of this relation in Fay's book is to show 
that the ratio of the higher-genus~$\Theta$-constant and a certain product of genus-1~$\theta$-functions (related to 
the right-hand side of~\eqref{Thth1th2reldiag}) is holomorphic as a function of the location of the branch points.
One can interpret this as a statement about the vanishing of the genus-2~$\Theta$-constant for the two-dimensional 
sub-moduli-space of genus-2 surfaces described by our branched coverings in the language of genus-1~$\theta$-functions.
In fact one can understand this vanishing for genus-2 in a very simple manner,  
this will become evident in the following section.

\section{Zeros of the~$\chi_t$ and~$\chi_g$, and their periodicity relations \label{sec:zeros}}

In this section we study the zeros of~$\chi_t$ and~$\chi_g$ and periodicity relations as a function of
the variable~$z_{12}$. In order to present a uniform treatment of the proof for all~$n$, it is convenient to 
define a new variable
\be
Z\= \begin{cases} 
\frac{z_{12}}{2n} & n \quad \text{even} \,,\\
\frac{z_{12}}{n} & n \quad \text{odd} \,. 
\end{cases} 
\label{defZ}
\ee
In the first subsection below we review the periodicity properties of~$\chi_g$ and~$\chi_t$ under shifts of the
variable~$Z$. We then argue that, given the periodicities, the knowledge of all  the zeros of these functions 
would be sufficient to prove the identity.  We then examine a subset of the zeros that are easy to identify. 
The details for even~$n$  and odd~$n$ turn out to be different, and so we treat them separately. 
For~$n=2$ the zeros we are able to identify are the only ones, which suffices to prove the identity in that case. 
In the following we will suppress the arguments corresponding to the spin structure~$\a,\b$ 
whenever there is no ambiguity, in order to avoid clutter. 

\subsection{Periodicity relations}

We start by recalling from~\cite{Mukhi:2017rex} the periodicity properties of the functions $\chi_{g}(\tau,z_{12};\a,\b)$ 
and~$\chi_{t}(\tau,z_{12};\a,\b)$ under translations of the argument~$z_{12}$. 
These are based on those of the $\th$-function with characteristics~$\a,\b \in (0,\half)$ \cite{Mumford:book}.
For~$\lambda, \mu \in \IZ$, 
we have the following two useful equations:\\
\ndt \emph{Integer shifts by lattice vectors:} 
\be
\label{thetaper}
\theta\bigg[\begin{matrix}\alpha\\ \beta\end{matrix}\bigg](z+\mu+\lambda\tau| \tau)
\=e^{2\pi i \alpha \mu} \, e^{-2\pi i \beta \lambda}e^{-i\pi \lambda^2\tau}e^{-2\pi i \lambda z}\,\theta\bigg[\begin{matrix}\alpha\\ \beta\end{matrix}\bigg](z| \tau) \,,
\ee
\emph{Half-shifts:}
\be
\label{thetaperhalf}
\begin{split}
\theta\bigg[\begin{matrix}\alpha\\ \beta\end{matrix}\bigg]\Bigl(z+\shalf
\big| \tau \Bigr)
&\=e^{2\pi i \alpha} \,\theta\bigg[\begin{matrix}\alpha\\ \beta-\half\end{matrix}\bigg](z| \tau) \,,\\
\theta\bigg[\begin{matrix}\alpha\\ \beta\end{matrix}\bigg]\Bigl(z+\shalf\tau
\big| \tau \Bigr)
&\=e^{-i\pi(z+\beta+\frac{\tau}{4})} \,\theta\bigg[\begin{matrix}\alpha-\half\\ \beta\end{matrix}\bigg](z| \tau) \,.
\end{split}
\ee
The periodicity relations of~$\chi_{t}$ follow immediately, with some differences between the case 
of even and odd~$n$ as we will now show.

\ndt {\bf Periodicity of $\chi_{t}$ for even~$n$}. The theta functions~$\theta\bigl(\frac{k}{n} z_{12}\big|\tau\bigr)$ 
appearing in the definition~\eqref{defchit} have $k \in \IZ + \half$.
In terms of~$Z=\frac{z_{12}}{2n}$, 
these theta functions are $\theta\bigl( 2k Z\big|\tau \bigr)$,
so that the arguments of all the theta functions in~\eqref{defchit} shift by integer multiples of~1 and~$\tau$ under shifts~$Z \to Z+1$ 
and~$Z \to Z+\tau$. Note that this is the smallest shift under which~$\chi_{t}(\t,z,\a,\b)$ is periodic, and it corresponds 
to a torus of sides~$(2n, 2n\t)$ in the variable~$z_{12}$. Using~\eqref{thetaper} we obtain:
\be
\label{chiperevenn}
\begin{split}
\chi_{t}(z_{12}+2n,\tau;\alpha,\beta) &\=  \chi_{t}(z_{12},\tau;\alpha,\beta) \,, \\
\chi_{t}(z_{12}+2n\tau,\tau;\alpha,\beta) &\=
e^{-i\pi\frac{n(n^2-1)}{3}\tau}e^{-i\pi\frac{n^2-1}{3} z_{12}}
\chi_{t}(z_{12},\tau;\alpha,\beta) \,.
\end{split}
\ee
There is also a periodicity relation obeyed by~$\chi_{t}(\t,z,\a,\b)$ under half-integer shifts of the argument~$Z$,
i.e.~$Z \to Z+\half$ and~$Z \to Z+\half \tau$. In this case the spin structures change according to Equation~\eqref{thetaperhalf}. 
We find:
\be
\label{chiperevennhalf}
\begin{split}
\chi_{t}(z_{12}+n,\tau;\alpha,\beta) & \= e^{2\pi i(\alpha-\half)}\chi_{t}(z_{12},\tau;\alpha,\beta-\shalf) \,, \\
\chi_{t}(z_{12}+n\tau,\tau;\alpha,\beta) & \=
e^{-\frac{i\pi n}{2}(\tau+ n(Z+\frac14))}
\chi_{t}(z_{12},\tau;\alpha-\shalf,\beta) \,.
\end{split}
\ee

\ndt {\bf Periodicity of $\chi_{t}$ for odd~$n$}. 
 The theta functions~$\theta\bigl(\frac{k}{n} z_{12}\big|\tau\bigr)$ 
appearing in~\eqref{defchit} now have $k \in \IZ$, 
which implies that the periodicity relations in~$z_{12}$ change compared to the even case. 
In terms of the variable~$Z=\frac{z_{12}}{n}$, they are $\theta\bigl(k Z\big|\tau \bigr)$, so that indeed 
under the shifts~$Z \to Z+1$ and~$Z \to Z+\tau$, the arguments of the theta functions again 
shift by integer multiples of~1 and~$\tau$.  Using~\eqref{thetaper} we have 
\be
\label{chiperoddn}
\begin{split}
\chi_{t}(z_{12}+n,\tau;\alpha,\beta) &\=\chi_{t}(z_{12},\tau;\alpha,\beta) \,, \\
\chi_{t}(z_{12}+n\tau,\tau;\alpha,\beta) &\=
e^{-i\pi\frac{n(n^2-1)}{12}\tau}e^{-i\pi\frac{n^2-1}{6} z_{12}}
\chi_{t}(z_{12},\tau;\alpha,\beta) \,.
\end{split}
\ee
Unlike the even case, there is no good periodicity property for~$\chi_t$ under the half-integer 
shifts~$Z \to Z+\half$ and~$Z \to Z+\half \tau$, as the arguments of the theta functions with odd~$k$ 
appearing in the definition~\eqref{defchit} change spin structure, while those with even~$k$ do not.

\vspace{0.4cm}
\ndt {\bf Periodicity of \bf $\chi_{g}$}. 
The periodicity relations for~$\chi_{g}$ do not follow immediately from their definition, as the variable~$z_{12}$
enters the~$\Theta$-function through the cut-differential~$\w_k$ and consequently through the period matrix~$\O$ 
(instead of as an elliptic variable). In fact there are two 
effects which need to be kept track of. First, the definition of the~$A$- and~$B$-cycles changes under the 
shifts of~$Z$ of the above type. Second, the period matrix itself changes by a corresponding shift. These effects 
were studied in Appendix A.1 of~\cite{Mukhi:2017rex} and the conclusion was that the higher-genus partition 
function~$\chi_g$ obeys the same periodicity relations as those of~$\chi_t$ for even~$n$
under both integer and half-integer shifts of~$Z$ (with the same change of spin structure in the latter case), 
and for odd~$n$ under integer shifts. This means equations~\eqref{chiperevenn}, \eqref{chiperevennhalf}, 
and~\eqref{chiperoddn} all hold if we replace~$\chi_t$ by~$\chi_g$ and $\alpha$, $\beta$ by $\valpha_{\rm diag}$, 
$\vbeta_{\rm diag}$, respectively.

\subsection{Zeros of~$\chi_g$ and~$\chi_t$ for even~$n$}

The significance of the zeros of~$\chi_g$ and~$\chi_t$ as functions of~$Z$ is the following.
Since both functions have the same 
periodicity properties under shifts of~$Z$ by integer multiples of~$1$ and~$\tau$, 
the ratio~$\chi_g/\chi_t$ is a well-defined function on the torus~$\IC/(\IZ\tau+\IZ)$. 
If we can show that~$\chi_g$ and~$\chi_t$ have the same zeros and poles in~$Z$, then the ratio~$\chi_g/\chi_t$ 
is a holomorphic function on the torus, and therefore a constant (by Liouville's theorem). 
In fact neither~$\chi_g$ nor~$\chi_t$ has poles in~$Z$, because theta functions are holomorphic in all their variables 
and the periods in the denominator of~\eqref{defchig} do not vanish.
In order to use the above argument, it is therefore enough to show that any zero of~$\chi_t$ is a zero of~$\chi_g$ 
of at least the same order, so that the ratio~$\chi_g/\chi_t$ would be a holomorphic function and hence constant. 
Once this is done, it is straightforward to evaluate the constant. 

Equation~\eqref{chiperevennhalf} shows that for even~$n$ the four different spin structures that we consider 
are related to each other by half-integer shifts of~$Z$. This means it is enough to study the zeros for any one 
of the spin-structures. We will see that~$(\half,\half)$ is the most convenient choice. 

We begin by recalling the zeros of~$\chi_t(z_{12},\t)$. Rewriting the definition~\eqref{defchit} in the variable~$Z$, we have: 
\be 
\chi_{t}\Bigl(\tau,z_{12}=2nZ;\shalf,\shalf \Bigr) \= 
\prod_{k=-{\frac{n-1}{2}}}^{\frac{n-1}{2}}\theta_{1} (2kZ | \tau) \,.
\ee
This function has a zero whenever any of its factors does. In the fundamental domain of~$Z$,~$\theta_{1}(2kZ)$ 
has simple zeros at~$Z=0$ and $Z=\frac{j}{2k}$,  $\frac{j}{2k} \tau$,  and $\frac{j}{2k} (\tau+1)$
with~$j=1,\cdots, 2k-1$. The corresponding factor~$\theta_{1}(-2kZ)$ has zeros at the same values. Thus 
the pattern of zeros of the entire expression $\chi_t$ is rather complicated, since zeros from different factors 
can occur at the same point. There are however two simple observations one can make: (i) for $n=2$, 
since $k=\pm\shalf$, there are no additional zeros beyond the double zero at the origin, (ii) for all other 
even $n$ there is an $n$th order zero at the origin, in addition to various other zeros in the fundamental domain of $Z$.

It is easy to calculate the coefficient of the~$n^\text{th}$-order zero at~$Z=0$, around which the expansion is:
\be \label{chitZ0}
\begin{split}
\chi_t \Bigl(z_{12}=2n Z,\t;\half,\half \Bigr)&\= \prod_{k=-\frac{n-1}{2}}^{\frac{n-1}{2}} \, k^2 \; \bigl( \theta'_1(0|\tau) \bigr)^n (2Z)^n 
+ \cO(Z^{n+1}) \, ,\\
&\= (-1)^{\frac{n}{2}} \bigl((n-1)!!\bigr)^2\,\bigl( \theta'_1(0|\tau) \bigr)^n Z^n  + \cO(Z^{n+1}) \, .
\end{split}
\ee

Now we turn to the higher genus expression $\chi_g$. The \Thth identity requires (and is implied by) 
the fact that all the zeros discussed above are also zeros of~$\chi_{g}$. 
Unfortunately it seems difficult to show that most of these are zeros of~$\chi_g$, but we can 
show it for the~$n^{\text{th}}$-order zero at~$Z=0$. For this, recall the cut differential and 
express it in terms of~$Z$:
\be
\omega_k (u,Z)\=\frac{\theta_1(u)}{\theta_1(u+2kZ)^{(1-\frac{k}{n})} \, \theta_1(u-2(n-k)Z)^{\frac{k}{n}}} \; ,\qquad k=0,1,\cdots, n-1 \, ,
\label{omkrecall}
\ee
where the $\tau$-dependence has been suppressed to simplify the notation. Notice that this is invariant 
under the simultaneous transformation $k\to n-k$ and $Z\to -Z$.
We expand this to second order in $Z$ and find:
\be
\omega_k(u,Z)\=1+ 2 \, k(k-n)\bigl(\log\theta_1(u)\bigr)'' Z^2+{\cal O}(Z^3) \,.
\ee
The term of order $Z$ vanishes. The next step is to compute the integrals:
\be \label{ABexp}
\begin{split}
A_{0k} (Z,\t)\; := \; \int_0^1 \omega_k\, du&\=1+ 2\, k(k-n) Z^2\int_0^1
\bigl(\log\theta_1(u)\bigr)'' du +{\cal O}(Z^3) \,, \\
&\= 1+{\cal O}(Z^3) \,, \\
B_{0k} (Z,\t)\; := \; \int_0^\tau \omega_k\, du&\=\tau+ 2\, k(k-n) Z^2\int_0^\tau
\bigl(\log\theta_1(u)\bigr)'' du +{\cal O}(Z^3) \,, \\
&\= \tau-i 4\pi k(k-n) Z^2+{\cal O}(Z^3) \,.
\end{split}
\ee
Here we used the identities:
\be
\begin{split}
\bigl(\log\theta_1\bigr)'(u+1)&\=\bigl(\log\theta_1\bigr)'(u) \,,\\
\bigl(\log\theta_1\bigr)'(u+\tau)&\=\bigl(\log\theta_1\bigr)'(u)-2\pi i \,.
\end{split}
\ee
Defining~$\Chat_k(Z,\tau)$ through  
\be
\frac{B_{0k}(Z,\t)}{A_{0k}(Z,\t)} \= \tau-i\pi \, \Chat_k(Z,\tau) \,,
\ee
it follows from~\eqref{ABexp} that:
\be
\Chat_k(Z,\tau) \= 4 \, k(k-n) Z^2+{\cal O}(Z^3) \,.
\ee
This enables us to rewrite the matrix $\Omega$ as follows:
\be
\begin{split}
\Omega_{ab}(Z,\t)&\=\frac{1}{n}\sum_{k=0}^{n-1}\cos\left(\frac{2\pi(a-b)k}{n}\right)\, \frac{B_{0k}(Z,\t)}{A_{0k}(Z,\t)}  \,,\\
&\=\tau\delta_{ab}-\frac{i\pi}{n} \sum_{k=0}^{n-1}\Chat_k(Z,\t) \,\cos\left(\frac{2\pi(a-b)k}{n}\right) \,.
\end{split}
\ee
We define the matrix:
\be
g_{ab}(Z,\tau) \= \frac{1}{2n}\sum_{k=0}^{n-1} \Chat_k (Z,\tau) \cos\left(\frac{2\pi (a-b) k}{n}\right) \,,\qquad
a,b=1,\cdots,n \,.
\ee
It is clear from this definition that $g_{ab}$ depends only on the difference $a-b$. Such a matrix is 
called a Toeplitz matrix. Now we can write:
\be\label{Omg}
\Omega_{ab}(Z,\t)\=\tau\delta_{ab} - 2 \pi i \, g_{ab}(Z,\tau)  \,.
\ee
The leading contribution to $g_{ab}$ is at order $Z^2$:
\be \label{gfrel}
g_{ab}(Z,\tau)\= Z^2 f_{ab}+{\cal O}(Z^3) \,.
\ee
where we have defined:
\be\label{fabdef}
\begin{split}
f_{ab} &\= \frac{2}{n}\sum_{k=0}^{n-1}k(k-n)\cos\left(\frac{2\pi(a-b)k}{n}\right) \\
&\=\frac{1}{\sin^2 \frac{\pi(a-b)}{n}} \, .
\end{split}
\ee
Like~$g_{ab}$,~$f_{ab}$ is also a Toeplitz matrix, and this point of view will become useful in a moment.

Now rewrite the function~$\Theta\bigg[\begin{matrix}\vec{\frac12}_\text{diag}\\ \vec{\frac12}_\text{diag}\end{matrix}\bigg]
\Bigl(0\big|\Omega(z_{12},\tau)\Bigr)$ using \eref{Omg}:
\be \label{ThetaExpDeriv}
\begin{split}
&\Theta\bigg[\begin{matrix}\vec{\frac12}_\text{diag}\\ \vec{\frac12}_\text{diag}\end{matrix}\bigg]
\Big(0\big|\Omega(z_{12},\tau)\Big) \\
& \qquad\=\sum_{m_a \in \IZ + \frac12}\exp \Bigl( -2\pi i \sum_{a,b=1}^n i \pi g_{ab}(Z,\tau)\, m_a \, m_b  \Bigr) \,
\exp\sum_{c=1}^n\left( i\pi \tau  m_c^2+\pi i m_c \right)  \,, \\
& \qquad\=\sum_{m_a \in \IZ + \frac12}\exp \Bigl(- \frac{1}{2} \sum_{a,b=1}^n  g_{ab}(Z,\tau)\, \p_a \, \p_b \Bigr) \,
\exp\sum_{c=1}^n\left( i\pi \tau  m_c^2+2\pi i m_c (z_c+\sfrac12)\right)  \Bigg|_{z_c=0}  \,,\\
& \qquad\=\exp \Bigl( - \frac{1}{2}  \sum_{a,b=1}^n g_{ab}(Z,\tau)\, \p_a \, \p_b \, \Bigr) \,
 \prod_{c=1}^n \theta_1(z_c|\tau) \, \Bigg|_{z_c=0} \,. 
  \end{split}
\ee
What we have done is to introduce fictitious variables $z_a$ and replace the integers~$m_am_b$ 
multiplying $g_{ab}$ by derivatives with respect to these variables, which are set to zero at the end. 
This allows us to re-express the higher-genus~$\Theta$ as an infinite series of derivatives of 
Jacobi~$\theta$-functions, as in the last line.
 
We can now expand the right-hand side as a power series in~$Z$ by expanding the exponential in the 
last line of~\eqref{ThetaExpDeriv}. Since~$\theta_1(z|\tau)$ is an odd function of~$z$, this expression 
is zero unless an odd number of derivatives hits each~$\theta_1(z|\tau)$.  
This means that the first non-zero term in the expansion of the exponential is the~$(\frac{n}{2})^\text{th}$ order term,
when there are~$\frac{n}{2}$ pairs of~$(a,b)$ with~$a\neq b$ from the derivatives and they are all different. 
Each such term occurs~$\frac{n}{2}!$ times, cancelling the factor $1/\frac{n}{2}!$ from expanding 
the exponential to this order. Recalling that~$g_{ab}(Z,\tau) = O(Z^2)$ for all~$a,b$, this also 
means that the smallest non-zero power  (=$n$) of~$Z$ is given by this term. Thus we obtain, using~\eqref{gfrel}:
\be\label{ThetaExpanded}
\Theta\bigg[\begin{matrix}\vec{\frac12}_\text{diag}\\ \vec{\frac12}_\text{diag}\end{matrix}\bigg]
\Bigl(0\big|\Omega(z_{12},\tau)\Bigr) \= (-1)^{\frac{n}{2}} Z^n\, \hbox{Hf}\big(f_{ab}\big) 
\bigl( \theta'_1(0|\tau) \bigr)^n + O(Z^{n+1}) \,,
\ee
where Hf indicates the ``Hafnian'' \cite{Caianiello:book} of an even-dimensional matrix, defined by:
\be
\hbox{Hf}(M_{ab})\=\sum_{\sigma\in C_n} \prod_{i=1}^{\frac{n}{2}} M_{\sigma(2i-1)\sigma(2i)} \, .
\ee
where the sum is over the set $C_n$ of canonical permutations in $S_n$, namely those that 
satisfy~$\sigma(2i-1)<\sigma(2i)$ for every $i$ and also $\sigma(2i-1)< \sigma(2j-1)$ for every $i<j$.
For example, for a $4\times 4$ matrix we have:
\be
\hbox{Hf}(M_{ab})\=M_{12}M_{34}+M_{13}M_{24}+M_{14}M_{23} \, .
\ee
Note that this definition implies that the Hafnian only depends on the upper triangular part of the 
matrix (excluding the diagonal).

Comparing with \eref{chitZ0}, we see that the behaviour near $Z=0$ is precisely the same for 
both~$\chi_t(z_{12},\t;\a,\b)$ and~$\chi_g(z_{12},\t;\a,\b)$ in the case of the spin structure~$(\a,\b)=(\half,\half)$, 
as long as the Hafnian  of the Toeplitz matrix~$f_{ab}$ obeys the combinatoric identity:
\be
\hbox{Hf}\left(\frac{1}{\sin^2 \frac{\pi(a-b)}{n}}\right) \= \bigl((n-1)!!\bigr)^{2} \, ,
\ee
which we have verified numerically to several orders.

\subsection{Elementary proof of the conjecture for~$n=2$}

Using the result of the previous subsection, we can give an elementary proof of the~$\Theta$\,-\,$\theta$ 
relation $\chi_{g} = \chi_{t}$ of Equation~\eqref{MainIdentity} in the case~$n=2$, for each of the four diagonal 
spin-structures~$\a,\b = 0, \half$. 

As we have seen, for the~$(\half, \half)$ spin structure~$\chi_t$ has a double zero at~$Z=0$ and no other 
zeros in the fundamental domain. We have also seen that~$\chi_g$ has a double zero at~$Z=0$ with the 
same coefficient, thus proving equality according to the argument above! We do not actually need to 
establish the absence of additional zeros for $\chi_g$: since it has the same periodicity as $\chi_t$ and 
no poles, it necessarily must have the same number of zeros. Finally we use the fact that for even $n$, 
one can use half-shifts to go to the other spin structures. Thus the \Thth relation is proved at $n=2$. 

A crucial feature of this proof was that at $n=2$ and for the spin structure~$(\half,\half)$,~$\chi_t$ has 
no additional zeros away from the origin (and likewise for the other spin structures,~$\chi_t$ has zeros 
only at a single point on the $Z$-torus). For $n>2$ there are additional zeros on the~$Z$-torus and this 
is the key reason why a complete proof is lacking in those cases.

\subsection{Zeros of~$\chi_g$ and~$\chi_t$ for odd~$n$}

For odd~$n$, the four different spin structures are not related to each other by half-integer 
shifts of~$Z$, in contrast to the case of even~$n$. The above methods can nevertheless be used 
to study the zeros of the~$(\half,\half)$ spin-structure. This problem, unfortunately, is a little 
degenerate because both sides vanish identically in this case. The higher-genus expression 
vanishes because the~$(\vec{\frac12}_\text{diag}, \vec{\frac12}_\text{diag})$ spin structure is 
an odd spin structure for odd~$n$, and so the corresponding~$\Theta$-constant vanishes. 
The twist-field expression vanishes because the product over~$k$ in~\eqref{defchit} now 
runs over integers (in contrast to half-integers in the even~$n$ case) and includes~$k=0$, 
so that one of the factors is the odd Jacobi theta constant which is identically zero.

If we can divide both sides of the expression by this vanishing Jacobi theta constant, then we 
can hope to make sense of the identity even for the odd spin structure. In order to do so, 
we deform the spin structure by a small amount (we recall that the characteristics of a theta 
function are actually real-valued).\footnote{Our original conjecture was proposed for
characteristics of order 2. However, the present result suggests that it could extend to real 
characteristics. This may be worth pursuing in the future.} 
The twist-field expression now becomes 
\be 
\chi_{t}\Bigl(\tau,z_{12}=nZ;\shalf\!+\ve,\shalf\!+\ve \Bigr) \= 
\prod_{k=-{\frac{n-1}{2}}}^{\frac{n-1}{2}}\theta\bigg[\,\begin{matrix}\half\!+\ve\\ \half\!+\ve\end{matrix}\, \bigg] (kZ | \tau) \,.
\ee
The term we have to be careful about is the one with~$k=0$ which identically vanishes at~$\ve=0$.  
In the other~$\theta$-functions we can safely take the limit~$\ve \to 0$, and in that limit their 
zeros in the fundamental domain of~$Z$ are at~$Z=0$ and~$Z=\frac{j}{k}$,  $\frac{j}{k} \tau$,  and $\frac{j}{k} (\tau+1)$, 
with~$j=1,\cdots, (k-1)$.

We can thus write the expansion near~$Z=0$:
\be \label{chitZ0odd}
\begin{split}
& \lim_{\ve \to 0}~  \chi_{t}\Bigl(\tau,z_{12}=nZ;\shalf\!+\ve,\shalf\!+\ve \Bigr) \Big/ 
\theta\bigg[\,\begin{matrix}\half\!+\ve\\ \half\!+\ve\end{matrix}\, \bigg] (Z | \tau)  \\
&\qquad \qquad \qquad \= \prod_{k=-\frac{n-1}{2}}^{\frac{n-1}{2}} \, k \; \bigl( \theta'_1(0|\tau) \bigr)^n Z^n 
+ \cO(Z^{n+1}) \, ,\\
&\qquad \qquad \qquad  \= (-1)^{\frac{n-1}{2}} \bigl(\tfrac{n-1}{2} \bigr)!^{\,2}\,\bigl( \theta'_1(0|\tau) \bigr)^n Z^n  + \cO(Z^{n+1}) \, .
\end{split}
\ee

Now we turn to the higher genus expression. The analysis for even~$n$ from~\eref{omkrecall} to \eref{fabdef} go 
through with the only change~$Z \to Z/2$ (because of the different definitions~\eqref{defZ}). Thus we reach:
\be \label{ThetaExpDerivOdd}
\begin{split}
&\Theta\bigg[\begin{matrix}(\vec{\half} +  \vec{\ve})_\text{diag}\\ (\vec{\half} +  \vec{\ve})_\text{diag}\end{matrix}\bigg]
\Big(0\big|\Omega(z_{12},\tau)\Big) \\
& \qquad\=\sum_{m_a \in \IZ + \frac12 + \ve}\exp \Bigl( -2\pi i \sum_{a,b=1}^n i \pi g_{ab}(Z,\tau)\, m_a \, m_b  \Bigr) \,
\exp\sum_{c=1}^n\Bigl( i\pi \tau  m_c^2+2\pi i m_c \bigl(\frac12 + \ve\bigr)\Bigr)  \,, \\
& \qquad\=\sum_{m_a \in \IZ + \frac12 + \ve}\exp \Bigl(- \frac{1}{2} \sum_{a,b=1}^n  g_{ab}(Z,\tau)\, \p_a \, \p_b \Bigr) \,
\exp\sum_{c=1}^n\Bigl( i\pi \tau  m_c^2+2\pi i m_c \bigl(z_c+\sfrac12 + \ve \bigr)\Bigr)  \Bigg|_{z_c=0}  \,,\\
& \qquad\=\exp \Bigl( - \frac{1}{2}  \sum_{a,b=1}^n g_{ab}(Z,\tau)\, \p_a \, \p_b \, \Bigr) \,
 \prod_{c=1}^n \theta\bigg[\,\begin{matrix}\half\!+\ve\\ \half\!+\ve\end{matrix}\, \bigg] (z_c | \tau) \Bigg|_{z_c=0} \,. 
  \end{split}
\ee

When we expand the right-hand side as a power series in~$Z$ as before, we see that this expression 
identically vanishes as~$\ve \to 0$, as at any order of the expansion there are an even number of derivatives hitting 
an odd number of~$\theta$-functions so that there is always at least one odd Jacobi theta function that is 
evaluated at~$z=0$. This also makes it clear that the limit
\be 
\lim_{\ve \to 0} \Biggr( \Theta\bigg[\begin{matrix}(\vec{\half} +  \vec{\ve})_\text{diag}\\ 
(\vec{\half} +  \vec{\ve})_\text{diag}\end{matrix}\bigg]
\Big(0\big|\Omega(z_{12},\tau)\Big)  \Big/ 
\theta\bigg[\,\begin{matrix}\half\!+\ve\\ \half\!+\ve\end{matrix}\, \bigg] (Z | \tau) \Biggr)
\ee
exists and is non-zero. The same type of combinatorics as in the even case now shows that the smallest 
non-vanishing order of~$Z$ appears from the term of order~$(n-1)/2$ which has~$n-1$ derivatives 
acting on~$n-1$ Jacobi~$\theta$-functions. There are now~$n$ ways to choose the~$n-1$ derivatives
(because one could omit any one of the~$z_c$). From the same type of calculation as before, we 
now obtain:
\be\label{ThetaExpandedOdd}
\begin{split}
\lim_{\ve \to 0} \Biggr( \Theta\bigg[\begin{matrix}(\vec{\half} +  \vec{\ve})_\text{diag}\\ 
(\vec{\half} +  \vec{\ve})_\text{diag}\end{matrix}\bigg]
\Big(0\big|\Omega(z_{12},\tau)\Big)  \Big/ 
\theta\bigg[\,\begin{matrix}\half\!+\ve\\ \half\!+\ve\end{matrix}\, \bigg] (Z | \tau) \Biggr)\\
\qquad \qquad \qquad \= (-1)^{\frac{n-1}{2}} Z^{n-1}\, \hbox{Hf}\big(f_{ab}\big) 
\bigl( \theta'_1(0|\tau) \bigr)^{n-1} + O(Z^{n}) \,,
\end{split}
\ee
where this time the matrix elements~$f$ are:
\be
f_{ab}  \=\frac{1}{4\sin^2 \frac{\pi (a-b)}{n}} \, ,
\ee
It is important to note that, although $f$ looks the same as in the even case, here the matrix itself is an~$(n-1) \times (n-1)$ matrix. Thus it is distinct from the $f$ matrix that appeared for the even case discussed in the previous section.
The corresponding combinatoric identity now is:
\be
\hbox{Hf}\left(\frac{1}{4\sin^2 \frac{\pi (a-b)}{n}}\right) \= \bigl(\tfrac{n-1}{2} \bigr)!^{\,2} \, , \qquad a,b=1,\cdots,n-1\, ,
\ee
which we have also verified to several orders.

\section{Concluding remarks \label{sec:conclusion}}

The primary motivation of this investigation was to understand the nature of the \Renyi entropy in a theory of free 
$1+1$-dimensional fermions
on a periodic space at finite temperature. This investigation led us to conjecture a non-trivial mathematical relation between 
higher-genus~$\Theta$-constants and genus one theta functions that follows from the fact that there are two methods  
(higher-genus and twist-field) of the same calculation 
which are mutually consistent. One could even regard these considerations as a physics proof of the relation using 
path-integral based methods for free fields and orbifolds. 

In this paper we presented two completely mathematical proofs of this relation for genus~$n=2$ and evidence 
(i.e.~checks of interesting consequences of the conjecture) for genus~$n >2$. From the mathematical point of view, 
we stumble upon the Schottky problem, that is to characterize the space of Jacobian varieties inside the space of abelian varieties, 
within the reduced context of cyclic~$n$-sheeted coverings of a genus-1 surface ramified at two points.

For the~$n=2$ case, the ideas of Fay gave a nice geometric characterization of this problem, which we used to solve it. 
The picture is that the Jacobian of the genus-2 surface splits into an image of the original torus and the Prym variety. 
The two pieces have eigenvalues~$\pm 1$ under the automorphism of the genus-2 surface induced by the covering map. 
The left-hand side of our identity~\eqref{MainIdentity} naturally comes from the Jacobian of the 2-sheeted cover 
(evaluated at the origin), while the right-hand side is naturally associated with the two pieces of this split. 

While we don't have a rigorous mathematical proof for~$n>2$, it seems to us 
that a generalization of the same idea should apply for any~$n$, although we are not aware of the 
analog of the theory of Prym varieties for higher covers. In this case, the Jacobian of the cyclic~$n$-sheeted 
cover splits into~$n$ pieces, each of whose eigenvalues are the~\nth roots of unity under an 
automorphism. The two sides of the conjecture~\eqref{MainIdentity} are now associated with these 
two geometric pictures, respectively.

\section*{Acknowledgements}

We would like to thank Nicholas Shepherd-Barron for interesting and useful discussions. 
The work of $\text{SM}_1$ was partially supported by a J.C.~Bose Fellowship, 
Government of India and that of $\text{SM}_2$  was supported by the ERC Consolidator 
Grant N.~681908, ``Quantum black holes: A macroscopic window into the microstructure of gravity'', 
and by the STFC grant ST/P000258/1. 
Parts of this work were done together at King's College London, at IISER Pune, and at ICTS Bengaluru.
$\text{SM}_1$ would like to thank King's College London and ICTS Bengaluru, and  
$\text{SM}_2$ would like to thank IISER Pune and ICTS Bengaluru, for their hospitality. 
%$\{\text{SM}_i\}$, $i=1,2$ 

%\bibliography{Allspinstructures}

\begin{thebibliography}{10}

\bibitem{Calabrese:2009qy}
P.~Calabrese and J.~Cardy, \emph{{Entanglement entropy and conformal field
  theory}},
  \href{https://doi.org/10.1088/1751-8113/42/50/504005}{\emph{J.Phys.}
  {\bfseries A42} (2009) 504005},
  [\href{https://arxiv.org/abs/0905.4013}{{\ttfamily 0905.4013}}].

\bibitem{Calabrese:2009ez}
P.~Calabrese, J.~Cardy and E.~Tonni, \emph{{Entanglement entropy of two
  disjoint intervals in conformal field theory}},
  \href{https://doi.org/10.1088/1742-5468/2009/11/P11001}{\emph{J.Stat.Mech.}
  {\bfseries 0911} (2009) P11001},
  [\href{https://arxiv.org/abs/0905.2069}{{\ttfamily 0905.2069}}].

\bibitem{Calabrese:2010he}
P.~Calabrese, J.~Cardy and E.~Tonni, \emph{{Entanglement entropy of two
  disjoint intervals in conformal field theory II}},
  \href{https://doi.org/10.1088/1742-5468/2011/01/P01021}{\emph{J.Stat.Mech.}
  {\bfseries 1101} (2011) P01021},
  [\href{https://arxiv.org/abs/1011.5482}{{\ttfamily 1011.5482}}].

\bibitem{Calabrese:2012ew}
P.~Calabrese, J.~Cardy and E.~Tonni, \emph{{Entanglement negativity in quantum
  field theory}},
  \href{https://doi.org/10.1103/PhysRevLett.109.130502}{\emph{Phys. Rev. Lett.}
  {\bfseries 109} (2012) 130502},
  [\href{https://arxiv.org/abs/1206.3092}{{\ttfamily 1206.3092}}].

\bibitem{Calabrese:2014yza}
P.~Calabrese, J.~Cardy and E.~Tonni, \emph{{Finite temperature entanglement
  negativity in conformal field theory}},
  \href{https://doi.org/10.1088/1751-8113/48/1/015006}{\emph{J. Phys.}
  {\bfseries A48} (2015) 015006},
  [\href{https://arxiv.org/abs/1408.3043}{{\ttfamily 1408.3043}}].

\bibitem{Azeyanagi:2007bj}
T.~Azeyanagi, T.~Nishioka and T.~Takayanagi, \emph{{Near Extremal Black Hole
  Entropy as Entanglement Entropy via AdS(2)/CFT(1)}},
  \href{https://doi.org/10.1103/PhysRevD.77.064005}{\emph{Phys.Rev.} {\bfseries
  D77} (2008) 064005}, [\href{https://arxiv.org/abs/0710.2956}{{\ttfamily
  0710.2956}}].

\bibitem{Datta:2013hba}
S.~Datta and J.~R. David, \emph{Renyi entropies of free bosons on the torus and
  holography}, \href{https://doi.org/10.1007/JHEP04(2014)081}{\emph{JHEP}
  {\bfseries 1404} (2014) 081},
  [\href{https://arxiv.org/abs/1311.1218}{{\ttfamily 1311.1218}}].

\bibitem{Cardy:2014jwa}
J.~Cardy and C.~P. Herzog, \emph{{Universal Thermal Corrections to Single
  Interval Entanglement Entropy for Two Dimensional Conformal Field Theories}},
  \href{https://doi.org/10.1103/PhysRevLett.112.171603}{\emph{Phys.Rev.Lett.}
  {\bfseries 112} (2014) 171603},
  [\href{https://arxiv.org/abs/1403.0578}{{\ttfamily 1403.0578}}].

\bibitem{Chen:2015cna}
B.~Chen and J.-q. Wu, \emph{{R{\'e}nyi entropy of a free compact boson on a
  torus}}, \href{https://doi.org/10.1103/PhysRevD.91.105013}{\emph{Phys. Rev.}
  {\bfseries D91} (2015) 105013},
  [\href{https://arxiv.org/abs/1501.00373}{{\ttfamily 1501.00373}}].

\bibitem{Lokhande:2015zma}
S.~F. Lokhande and S.~Mukhi, \emph{{Modular invariance and entanglement
  entropy}}, \href{https://doi.org/10.1007/JHEP06(2015)106}{\emph{JHEP}
  {\bfseries 06} (2015) 106},
  [\href{https://arxiv.org/abs/1504.01921}{{\ttfamily 1504.01921}}].

\bibitem{Mukhi:2017rex}
S.~Mukhi, S.~Murthy and J.-Q. Wu, \emph{{Entanglement, Replicas, and Thetas}},
  \href{https://doi.org/10.1007/JHEP01(2018)005}{\emph{JHEP} {\bfseries 01}
  (2018) 005}, [\href{https://arxiv.org/abs/1706.09426}{{\ttfamily
  1706.09426}}].

\bibitem{Fay:book}
J.~D. Fay, \emph{Theta Functions on Riemann Surfaces}.
\newblock Springer-Verlag, 1973.

\bibitem{AlvarezGaume:1986es}
L.~Alvarez-Gaume, G.~W. Moore and C.~Vafa, \emph{{Theta Functions, Modular
  Invariance and Strings}},
  \href{https://doi.org/10.1007/BF01210925}{\emph{Commun. Math. Phys.}
  {\bfseries 106} (1986) 1--40}.

\bibitem{Dijkgraaf:1987vp}
R.~Dijkgraaf, E.~P. Verlinde and H.~L. Verlinde, \emph{{C = 1 Conformal Field
  Theories on Riemann Surfaces}},
  \href{https://doi.org/10.1007/BF01224132}{\emph{Commun. Math. Phys.}
  {\bfseries 115} (1988) 649--690}.

\bibitem{Mumford:book}
D.~Mumford, \emph{Tata Lectures on Theta I}.
\newblock Birkh{\"a}user Basel, 1983.

\bibitem{Caianiello:book}
E.~Caianiello, \emph{Combinatorics and Renormalisation in Quantum Field
  Theory}.
\newblock W.A. Benjamin, 1973.

\end{thebibliography}
%\bibliographystyle{JHEP}

\providecommand{\href}[2]{#2}\begingroup\raggedright\endgroup

\end{document}